\documentclass[11pt,twoside]{article}
\usepackage{asp2010}

\resetcounters

\begin{document}

\title{Nearby Motionless Stars}
\author{Adric~R.~Riedel$^1$, Todd~J.~Henry$^1$, Russel~J.~White$^2$, Inseok~Song$^3$, Eric~L.~N.~Jensen$^4$, Nigel~C.~Hambly$^5$}
\affil{$^1$Georgia State University, Department of Physics and Astronomy, 29 Peachtree Center Avenue, Atlanta, GA 30302, USA/REsearch Consortium On Nearby Stars}
\affil{$^2$Georgia State University, Department of Physics and Astronomy, 29 Peachtree Center Avenue, Atlanta, GA 30302, USA}
\affil{$^3$Department of Physics and Astronomy, The University of Georgia, Athens, GA 30602-2451a, USA}
\affil{$^4$Department of Physics and Astronomy, 500 College Ave., Swarthmore College, Swarthmore, PA 19081, USA}
\affil{$^5$Institute for Astronomy, University of Edinburgh, Blackford Hill, Edinburgh, EH9 3HJ, Scotland, United Kingdom/Scottish Universities Physics Alliance}

\begin{abstract}

We present methods and preliminary results of a relatively novel
search for nearby stars.  The method relies on photometric distance
estimates as its primary search criterion, thus distinguishing itself
from proper motion-based searches that have produced the bulk of
nearby star discoveries.

\end{abstract}

\section{History of nearby star discoveries}
 Over the last 200 years, proper motion (the apparent motion of stars
 across the sky, seen over periods of years) has been used to find
 nearby stars.  This approach has been based on the idea that stars
 move through space, and the closest ones should appear to move the
 fastest.  Currently, we understand that this is due to the
 combination of the motions of the Sun and the star in question in
 their orbits around the center of the Galaxy, though this idea
 predates the discovery of the Galaxy and our place in it by at least
 150 years. The earliest reference available seems to be William
 Herschel \citep{1783RSPT...73..247H}, who claims the phenomena is
 well-established and credits its discovery to Sir Edmund Halley.

 This property of large proper motion has served nearby star research
 well from the very beginning, forming at least part of the decisions
 of \citet{1838MNRAS...4..152B} and \citet{1839MNRAS...4..168H} to
 observe 61 Cygni and Alpha Centauri (respectively) for parallax.  The
 trend continues to the present.  Nearly all nearby stars known have
 high proper motions, with the limit of high proper motion (0.5\arcsec
 yr$^{-1}$, van Maanen) or interesting proper motion (0.2\arcsec
 yr$^{-1}$, the Royal Greenwich Observatory) set by influential
 publications in the early part of the 20th century
 \citep{1988IAUS..133..301L}.  Though Luyten does not give the
 reasoning behind either argument, we suspect the 0.2\arcsec
 yr$^{-1}$~limit was likely the best that could be done visually with
 the photographic plates and blink comparators of the time.

 Still, there are signs proper motion is not an entirely foolproof
 method.  Proxima Centauri, the closest star to Earth, is only the
 18th highest proper motion star in the New Luyten Two Tenths Catalog
 \citep[][, though the other 17 are also very nearby
   stars]{1979nltt.book.....L} (though the other 17 are also very
 nearby stars).  By the same token, many brighter stars were singled
 out for parallax, e.g. Hipparcos observed all stars brighter than
 $V$=7.3 \citep{1997A&A...323L..49P} and discovered not all objects
 within 25 parsecs are fast moving.  One good example is Gl 566, which
 despite a distance of 6.7 parsecs is moving at 0.169\arcsec
 yr$^{-1}$.  It would likely not have been noticed except that it is a
 5th magnitude G star with a K companion and visible orbital motion.

 In any case, years later we still have unfinished business.  The
 situation is currently thus: assuming that 50 parallax-verified
 systems within 5 parsecs is a statistically meaningful sample, there
 should be 400 stars within 10 parsecs (8 times the volume), and 6250
 systems within 25 parsecs (125 times the volume). The current tally
 is 256 parallax-verified systems within 10 parsecs\footnote{RECONS 10
   pc census, http://www.recons.org/census.posted.htm; Henry,
   T.J. accessed 2010-12-01} and 2011 parallax-verified systems within
 25 parsecs \citep[NStars,][]{2002AJ....123.2002H}. We are missing
 nearly 36\% of systems within 10 pc and 68\% of all stars within 25
 pc.

\section{Where are the stars?}

 The most obvious reason stars are missing is that we simply do not
 have accurate trigonometric parallaxes for all potential nearby stars
 we know of.  The Luyten Half Second catalog (2nd ed)
 \citep{1979lccs.book.....L}, New Luyten Two Tenths catalog
 \citep{1979nltt.book.....L}, and Giclas survey papers \citep[final
   entries,][]{1979LowOB...8..145G} contain the most complete samples
 of potentially nearby proper motion objects with over 50,000 stars,
 and were published three decades ago.  We still have no reliable
 trigonometric parallax for large numbers of these high-proper-motion
 systems \citep[e.g.][]{2010AJ....140..897R}, particularly the ones
 fainter than the completeness limit of Hipparcos.  This is an area
 where existing parallax programs and future programs like LSST, URAT,
 Skymapper, PanSTARRS and Gaia will have a huge impact.

 There are still, however, many nearby stars we have not identified
 yet.  They fall into three basic groups:

 {\bf Stars simply missed by Luyten and Giclas.} Luyten's NLTT survey
 was done with hand-blinked plates in the far south from his earlier
 BPM survey (which had a limiting magnitude of $R$=16.5) and
 machine-scanned plates from the Luyten-Palomar survey ($R$=19) for
 northern regions, with a limit of $\mu >$ 0.18\arcsec yr$^{-1}$~to
 make certain he found all stars moving faster than 0.20\arcsec
 yr$^{-1}$.  Giclas' Lowell Proper Motion survey ($R$=16.5) had a
 nominal limit of 0.27\arcsec yr$^{-1}$~and a goal of finding every
 object moving faster than 0.30\arcsec yr$^{-1}$; that proper motion
 limit was later reduced to 0.20\arcsec yr$^{-1}$~in the southern
 hemisphere, but the survey was left unfinished in 1979.
 \citet{2005AJ....129.1483L} estimate Luyten's $\sim$90\% completion
 limit to be $V$=15 within 10 degrees of the galactic plane, and
 $V$=18 elsewhere.  Many astronomers (too numerous to note) have had
 successful programs locating such objects, particularly in the south,
 where Giclas was unfinished and Luyten hand-blinked plates with
 brighter limiting magnitudes.

 { \bf Stars too faint to be seen by Luyten or Giclas.} Luyten and
 Giclas were both limited by the sensitivity and waveband of their
 first or second epoch plates. The smallest stars (M9.5V or L0V) have
 absolute magnitudes of $R$=18, therefore the Luyten and Giclas
 surveys cannot complete a nearby star sample out to a distance of 25
 pc (distance modulus $\sim$2), where such stars are roughly $R$=20.
 Surveys with fainter magnitude limits (including those that scan
 photographic plates to lower magnitude limits) and that use infrared
 wavebands are better suited to detect faint nearby stars.

 {\bf Stars moving too slowly for Luyten or Giclas.}  The limits set
 by Luyten (and the Royal Greenwich Observatory before him) were more
 out of necessity than scientific reason; going to smaller motions
 releases a flood of new objects that are much harder to measure
 accurately.  With the development of modern computing capabilities,
 though, such things become possible.  Many recent surveys have
 breached the 0.18\arcsec yr$^{-1}$~limit, although none have
 explicitly surveyed tiny proper motion objects.

\section{A different method for finding nearby stars: TINYMO}

\begin{figure}[!ht]
\plotone{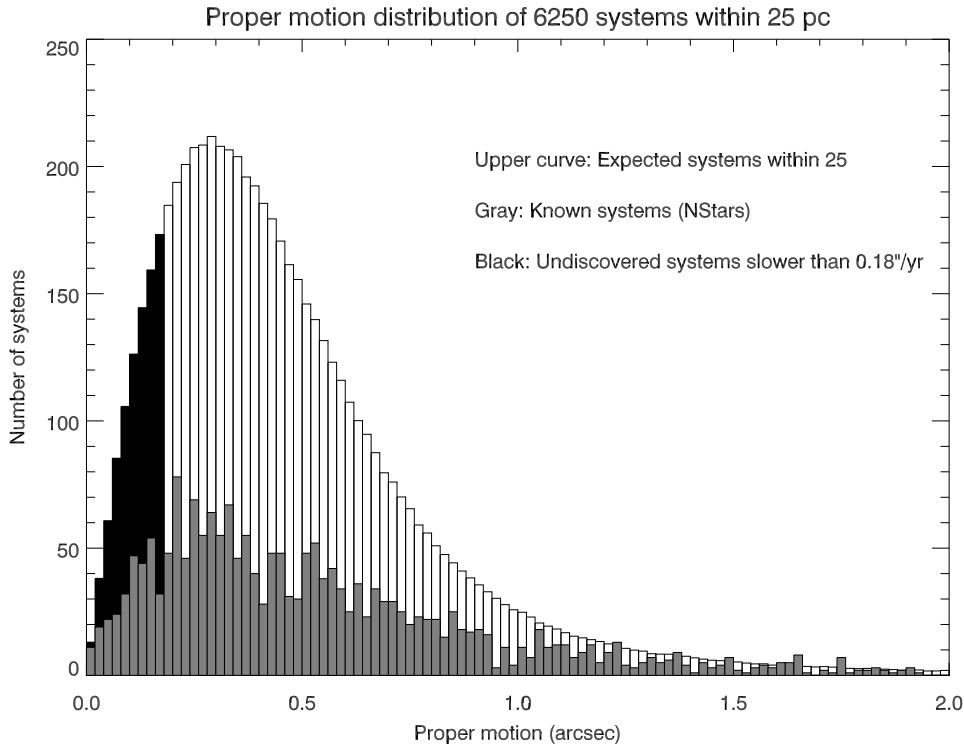}
\caption{A Monte-Carlo simulation of stars within 25 pc of the Sun
  (black/white) using velocity dispersions from
  \citet{2009MNRAS.397.1286A}; known objects (gray) are from the
  NStars 25 pc database
  \citep{2002AJ....123.2002H} \label{riedel_a_fig1}}
\end{figure}

We have chosen to focus our survey on finding extremely low proper
motion objects, rather than fainter stars.  Space velocity dispersions
calculated from known nearby stars (themselves potentially biased
toward high transverse velocities) can be used in Monte Carlo
simulations of nearby stars; one such simulation is shown in Figure
\ref{riedel_a_fig1}, where the distribution of stars within 25 pc is
modeled based on the velocity dispersions in
\citet{2009MNRAS.397.1286A} and color/number distributions from
RECONS\footnote{http://www.recons.org/census.posted.htm; Henry,
  T.J. accessed 2010-12-01}, assuming an actual population of 6250
systems.  As is visible in the figure, while most systems are moving
faster than 0.18\arcsec yr$^{-1}$, a sizable portion (14.6\% in this
simulation) are moving more slowly; only 31\% of those are
currently known. These undiscovered systems may be quite bright and
easy to study with current instrumentation.

The TINYMO survey is designed to detect these slow-moving nearby stars
using photometric distance estimates as the primary selection
criterion, avoiding (except for a rough upper limit) proper motions
entirely.  It thus avoids many of the pitfalls of using proper motion
to search for truly zero proper motion\footnote{Interestingly, there
  is also a problem with detecting high proper motion objects, which
  \citet{2004AJ....128.2460H} and \citet{2005ApJ...633L.121L} prove
  are still missing; for the most part the problem is identifying
  widely separated ``single'' hits as an individual moving star,
  rather than plate defects or transient sources} nearby stars, most
notably a.) there is a finite limit to how well proper motions may be
measured from any given source, and b.) even distant background
objects are moving at some level.

The so-called TINYMO search for essentially non-moving stars was
conducted (much like previous RECONS searches in
\citet{2005AJ....129..413S} and \citet{2007AJ....133.2898F}) using the
SuperCOSMOS Sky Survey \citep{2001MNRAS.326.1279H} to search for tiny
proper motion stars in the southern hemisphere, among the $\sim$1
billion distinct catalog entries.  The empirical SuperCOSMOS $B_JR_2I$
and 2MASS $JHK_s$ plate photometry relations in
\citep{2004AJ....128..437H} were used to obtain distance estimates to
these targets, with the intent to select stars within 25 pc.

Spatially, TINYMO (currently) only contains stars in the southern
hemisphere, more than 10 degrees from the galactic plane and more than
20 degrees from the galactic bulge.  This time, we required targets to
be detected on all four ($B_J$, $R_1$, $R_2$, $I$\footnote{$B_J$ is
  the Science and Engineering Research Council (SERC)-J survey; $R_1$
  is POSS-I E between 0 DEC and -18 DEC and ESO-R below -20 DEC; $R_2$
  is SERC-ER$_{59F}$, and $I$ is SERC-I$_{IVN}$}) plates \citep[which
  effectively required all four detections to be within 6\arcsec of
  each other,][]{2001MNRAS.326.1279H} and with a 2MASS detection
within 5\arcsec of the weighted mean plate position.  While that is
technically a proper motion limit, it is an {\it upper} limit,
dependent on the epoch spread of individual plates; it is merely
included to enforce the selection of only the slow-moving objects we
wish to study.

The nearly 14 million resulting non-moving objects were then run
through the SuperCOSMOS $B_JR_2I$ and 2MASS $JHK_s$ main-sequence
photometric color relations from \citet{2004AJ....128..437H}; less
than 89000 objects were estimated to be within 25 pc.

\begin{figure}[!ht]
\plotone{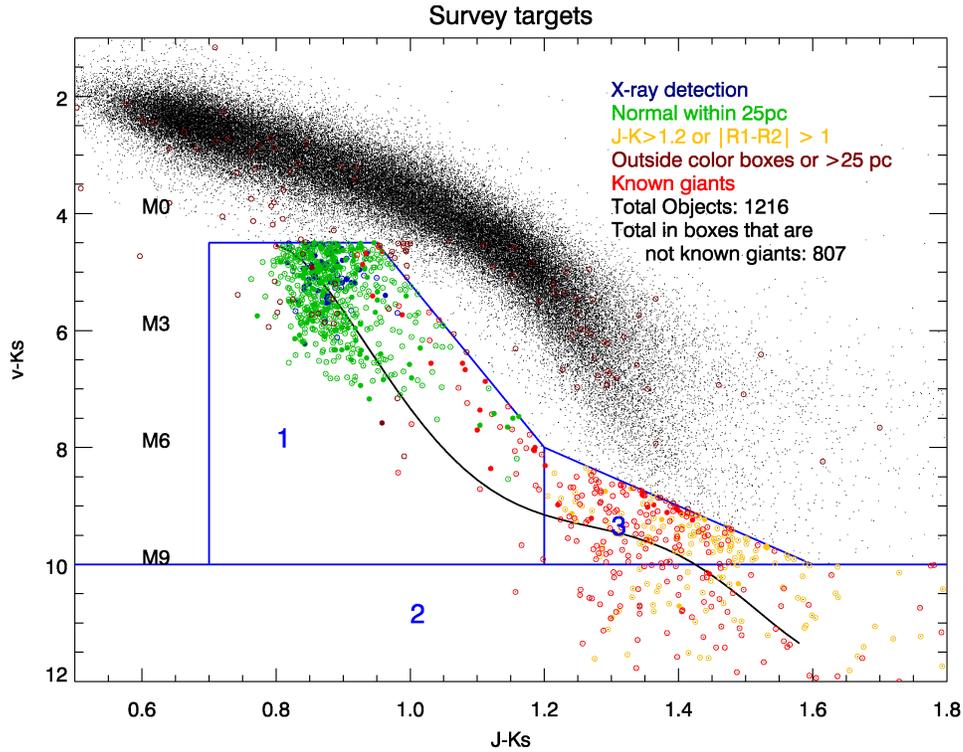}
\caption{The color-color selection boxes used for the TINYMO search,
  in $J$--$K_s$ vs $v$--$K_s$ space.  Curve is a fifth-order fit to
  the main sequence.  Filled circles have been followed up with CCD
  photometry.\label{riedel_a_fig2}}
\end{figure}

Finally, a color-color criterion in $J$-$K_s$ vs $v$-$K$ colorspace
(the average of SuperCOSMOS $B_J$ and $R_2$ is taken as our simulated
$v$) was applied, with the goal of separating main sequence objects
from giants.  The color-selection boxes can be seen in Figure
\ref{riedel_a_fig2}, plotted in $v$-$K_s$ vs $J$-$K_s$ space. A
decision was made early on to keep targets in boxes 2 and 3 (brown
dwarf colors), even though they are severely contaminated by giants.

At this stage, the 1154 remaining color-selected objects were blinked
using SuperCOSMOS plates and Aladin's SIMBAD overlay loaded into the Aladin
Skyview Applet, and individually examined to check if they were a.)
real objects, b.) moving (if possible), c.)  matched to the proper
2MASS point (mistakes in the 2MASS identification account for the open
circled points now lying outside the color-color boxes in Figure
\ref{riedel_a_fig2}), and d.) previously known objects.  At this
stage, several proper motion objects (usually companions) were
non-exhaustively identified by eye for a total of 1216 objects.
Proper motions (which did not form a selection criterion, though they
can still be calculated) for these objects range from 0.000\arcsec
yr$^{-1}$~to 0.444\arcsec yr$^{-1}$.

\section{Contamination, and how to deal with it}

We have thus far assumed every object in the SuperCOSMOS database is a
single main-sequence star.  This is not accurate, and leads to an
enormous contamination problem.  In particular, apart from subdwarfs
and (theoretically) white dwarfs, contaminants with the colors of
main-sequence stars are much brighter; they will have scattered INTO
the sample. The most common culprits are giants, especially Mira Ceti
variables and AGB stars, whose red colors can look very much like M
dwarfs.  To this we can also add carbon stars, distant objects
reddened by the ISM or a molecular cloud, multiple stars, young stars
and (again, theoretically)\footnote{SIMBAD did identify one X-ray
  bright object as a possible BL Lac; it is a star at $\sim$16 pc}
AGN.

These problems can be solved with additional color criteria, for
instance if the SuperCOSMOS $R_1$ and $R_2$ magnitudes do not match to
within a magnitude, or fewer than 9 of 11 photometric distance
relations produced valid results.  We have also done literature
searches to identify stars, particularly in the The General Catalog of
Variable Stars \citep[][in VizieR as b/GCVS]{2009yCat....102025S}
which maintains a list of all known variable stars and can be used to
identify Mira variables, Carbon stars, and other semi-regular and
irregular giant stars.

Finally, for stars we anticipate following up for parallax (i.e., $\mu
<$ 0.18\arcsec yr$^{-1}$~and plate distance within 15 pc -- thesis
timescales are short), we are obtaining $V_JR_{KC}I_{KC}$ CCD
photometry at CTIO, and low-resolution red (6000-9000\AA~at
10\AA~resolution) spectra on the CTIO 1.5m telescope with RC
Spectrograph, and the Lowell Observatory 1.8m telescope with the
DeVeny Spectrograph.  Stars that are still within 15 pc by the
$V_JR_{KC}I_{KC}JHK_s$ color-distance relations in
\citet{2004AJ....128.2460H} AND confirmed to not be giants are being
observed for parallax on the Cerro Tololo Inter-american Observatory
Parallax Investigation (CTIOPI) \citep[][and
  subsequent]{2005AJ....129.1954J} program.

\section{Results}

\begin{figure}[!ht]
\plotone{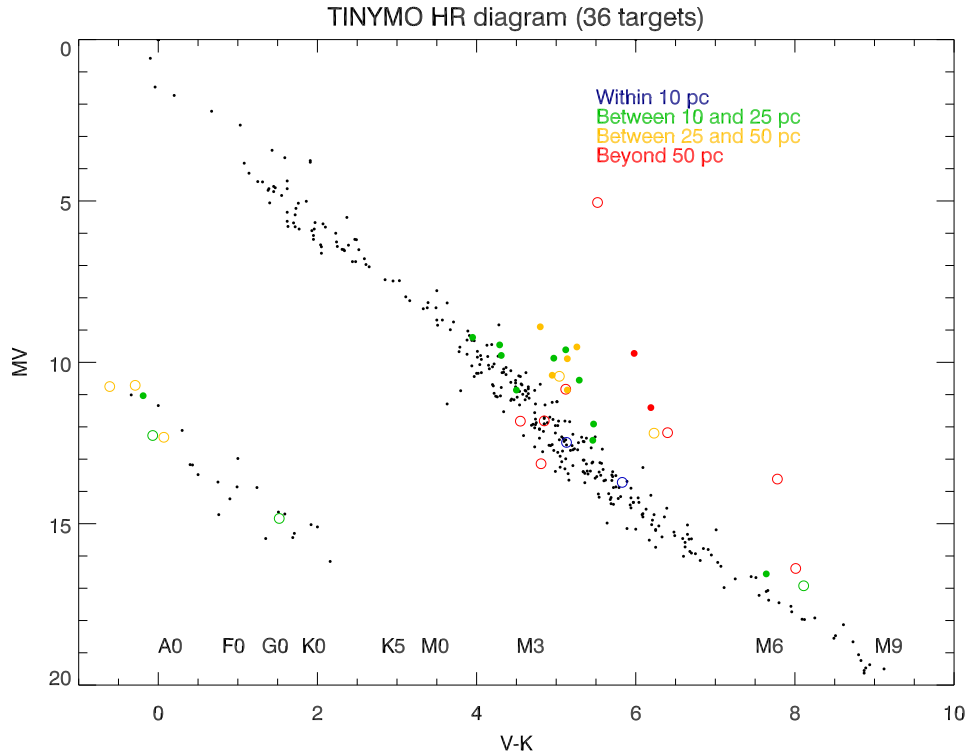}
\caption{Color-magnitude diagram showing the locations of 36 tiny
  proper motion objects in 32 systems with preliminary parallaxes from
  the CTIOPI program.  Filled circles are X-ray bright, small points
  are the RECONS 10 pc sample \label{riedel_a_fig3}}
\end{figure}

\begin{figure}[!ht]
\plotone{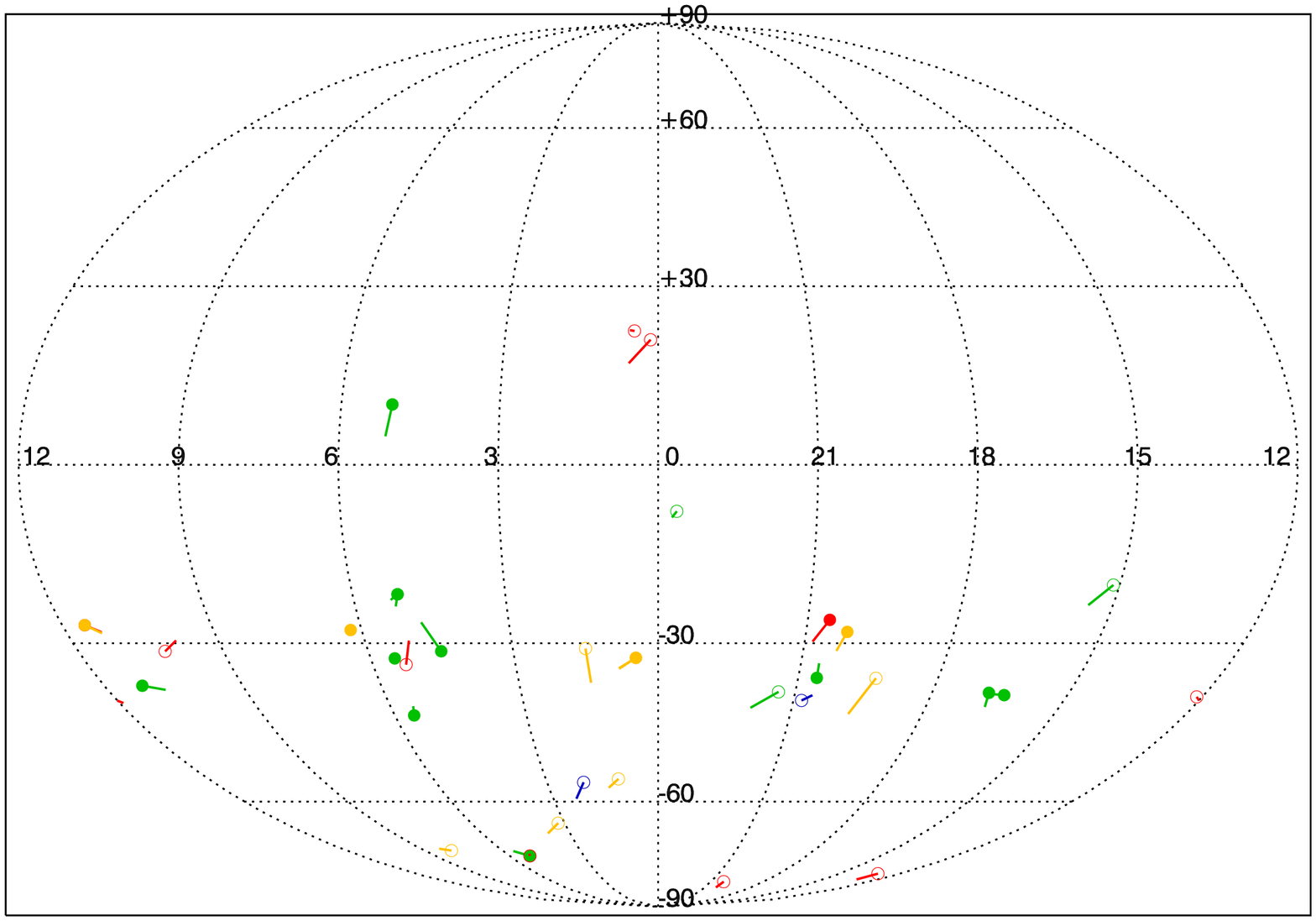}
\caption{Spatial plot of 36 tiny proper motion objects with parallaxes
  from the CTIOPI project.  Proper motion vectors have been scaled by
  180000.  Color and fill follow Figure
  \ref{riedel_a_fig3}. \label{riedel_a_fig4}}
\end{figure}

Including star systems like SCR2049-4012 at 9.2 pc and proper motion
0.06\arcsec yr$^{-1}$, we have preliminary parallax results for 36
stars in 32 star systems with proper motions less than 0.18\arcsec
yr$^{-1}$~from the CTIOPI program, consisting of targets from the
TINYMO search and other various additions. Sixteen of the parallax
targets are within 25 pc and 11 more are between 25 and 50 pc.  They
are plotted on a color-magnitude diagram in Figure
\ref{riedel_a_fig3}, and on the sky with transverse motion vectors in
Figure \ref{riedel_a_fig4}.

As can be seen in Figure \ref{riedel_a_fig3}, the vast majority are
either multiple, young, or both, lying above (in some cases, well above) the
main sequence.  Nearly the entire sample spectroscopically observed
thus far has H-alpha emission, also suggesting chromospheric activity
from either (relative) youth or close duplicity. At least some of this
is an observational bias in the sample toward objects with substantial
X-ray flux as measured in the ROSAT All-Sky surveys
\citep{1999A&A...349..389V,2000IAUC.7432R...1V}; early on it was
discovered that nearly every object with an IRAS detection was a
giant, and nearly everything with an X-ray detection was a genuine
nearby star with a chance of being young.  Accordingly, we have
focused on obtaining CCD photometry for every X-ray bright system
within 25 pc by plate photometry (rather than the usual 15),
occasionally the resulting CCD distance estimate is within 15 pc (by
$V_JR_{KC}I_{KC}$ CCD photometric distance) and the star is added to
the parallax program.

TINYMO (and the CTIOPI parallax program) reveals known members of Beta
Pic, TW Hydra and the highly reddening Chameleon I cloud.  Of
particular interest in Figure \ref{riedel_a_fig4} are three roughly
co-moving points clustered around 12h RA, -35 DEC corresponding to
three known TW Hydra members (including 2MASS 1207-3932=TWA 27,
\citealt{2007ApJ...669L..45G}). Three (maybe four) co-moving objects
clustered around 21h RA, -35 DEC are new discoveries that may belong
to another moving group.

\bibliography{riedel_a}

\begin{thebibliography}{}
\expandafter\ifx\csname natexlab\endcsname\relax\def\natexlab#1{#1}\fi
\expandafter\ifx\csname url\endcsname\relax
  \def\url#1{\texttt{#1}}\fi
\expandafter\ifx\csname urlprefix\endcsname\relax\def\urlprefix{URL }\fi
\providecommand{\eprint}[2][]{\url{#2}}

\bibitem[{{Aumer} \& {Binney}(2009)}]{2009MNRAS.397.1286A}
{Aumer}, M., \& {Binney}, J.~J. 2009, \mnras, 397, 1286. \eprint{0905.2512}

\bibitem[{{Bessel}(1838)}]{1838MNRAS...4..152B}
{Bessel}, F.~W. 1838, \mnras, 4, 152

\bibitem[{{Finch} et~al.(2007){Finch}, {Henry}, {Subasavage}, {Jao}, \&
  {Hambly}}]{2007AJ....133.2898F}
{Finch}, C.~T., {Henry}, T.~J., {Subasavage}, J.~P., {Jao}, W., \& {Hambly},
  N.~C. 2007, \aj, 133, 2898. \eprint{arXiv:astro-ph/0703133}

\bibitem[{{Giclas} et~al.(1979){Giclas}, {Burnham}, \&
  {Thomas}}]{1979LowOB...8..145G}
{Giclas}, H.~L., {Burnham}, R., Jr., \& {Thomas}, N.~G. 1979, Lowell
  Observatory Bulletin, 8, 145

\bibitem[{{Gizis} et~al.(2007){Gizis}, {Jao}, {Subasavage}, \&
  {Henry}}]{2007ApJ...669L..45G}
{Gizis}, J.~E., {Jao}, W., {Subasavage}, J.~P., \& {Henry}, T.~J. 2007, \apjl,
  669, L45. \eprint{0709.1178}

\bibitem[{{Hambly} et~al.(2004){Hambly}, {Henry}, {Subasavage}, {Brown}, \&
  {Jao}}]{2004AJ....128..437H}
{Hambly}, N.~C., {Henry}, T.~J., {Subasavage}, J.~P., {Brown}, M.~A., \& {Jao},
  W. 2004, \aj, 128, 437. \eprint{arXiv:astro-ph/0404265}

\bibitem[{{Hambly} et~al.(2001){Hambly}, {MacGillivray}, {Read}, {Tritton},
  {Thomson}, {Kelly}, {Morgan}, {Smith}, {Driver}, {Williamson}, {Parker},
  {Hawkins}, {Williams}, \& {Lawrence}}]{2001MNRAS.326.1279H}
{Hambly}, N.~C., {MacGillivray}, H.~T., {Read}, M.~A., {Tritton}, S.~B.,
  {Thomson}, E.~B., {Kelly}, B.~D., {Morgan}, D.~H., {Smith}, R.~E., {Driver},
  S.~P., {Williamson}, J., {Parker}, Q.~A., {Hawkins}, M.~R.~S., {Williams},
  P.~M., \& {Lawrence}, A. 2001, \mnras, 326, 1279.
  \eprint{arXiv:astro-ph/0108286}

\bibitem[{{Henderson}(1839)}]{1839MNRAS...4..168H}
{Henderson}, T. 1839, \mnras, 4, 168

\bibitem[{{Henry} et~al.(2004){Henry}, {Subasavage}, {Brown}, {Beaulieu},
  {Jao}, \& {Hambly}}]{2004AJ....128.2460H}
{Henry}, T.~J., {Subasavage}, J.~P., {Brown}, M.~A., {Beaulieu}, T.~D., {Jao},
  W., \& {Hambly}, N.~C. 2004, \aj, 128, 2460. \eprint{arXiv:astro-ph/0408240}

\bibitem[{{Henry} et~al.(2002){Henry}, {Walkowicz}, {Barto}, \&
  {Golimowski}}]{2002AJ....123.2002H}
{Henry}, T.~J., {Walkowicz}, L.~M., {Barto}, T.~C., \& {Golimowski}, D.~A.
  2002, \aj, 123, 2002. \eprint{arXiv:astro-ph/0112496}

\bibitem[{{Herschel}(1783)}]{1783RSPT...73..247H}
{Herschel}, W. 1783, Royal Society of London Philosophical Transactions Series
  I, 73, 247

\bibitem[{{Jao} et~al.(2005){Jao}, {Henry}, {Subasavage}, {Brown}, {Ianna},
  {Bartlett}, {Costa}, \& {M{\'e}ndez}}]{2005AJ....129.1954J}
{Jao}, W., {Henry}, T.~J., {Subasavage}, J.~P., {Brown}, M.~A., {Ianna}, P.~A.,
  {Bartlett}, J.~L., {Costa}, E., \& {M{\'e}ndez}, R.~A. 2005, \aj, 129, 1954.
  \eprint{arXiv:astro-ph/0502167}

\bibitem[{{L{\'e}pine} et~al.(2005){L{\'e}pine}, {Rich}, \&
  {Shara}}]{2005ApJ...633L.121L}
{L{\'e}pine}, S., {Rich}, R.~M., \& {Shara}, M.~M. 2005, \apjl, 633, L121.
  \eprint{arXiv:astro-ph/0510101}

\bibitem[{{L{\'e}pine} \& {Shara}(2005)}]{2005AJ....129.1483L}
{L{\'e}pine}, S., \& {Shara}, M.~M. 2005, \aj, 129, 1483.
  \eprint{arXiv:astro-ph/0412070}

\bibitem[{{Luyten}(1979{\natexlab{a}})}]{1979lccs.book.....L}
{Luyten}, W.~J. 1979{\natexlab{a}}, {LHS catalogue. A catalogue of stars with
  proper motions exceeding 0''5 annually}

\bibitem[{{Luyten}(1979{\natexlab{b}})}]{1979nltt.book.....L}
--- 1979{\natexlab{b}}, {NLTT catalogue. Volume\_I. +90\_\_to\_+30\_.
  Volume.\_II. +30\_\_to\_0\_.}

\bibitem[{{Luyten}(1988)}]{1988IAUS..133..301L}
--- 1988, in Mapping the Sky: Past Heritage and Future Directions, edited by
  {S.~Debarbat}, vol. 133 of IAU Symposium, 301

\bibitem[{{Perryman} et~al.(1997){Perryman}, {Lindegren}, {Kovalevsky}, {Hoeg},
  {Bastian}, {Bernacca}, {Cr{\'e}z{\'e}}, {Donati}, {Grenon}, {van Leeuwen},
  {van der Marel}, {Mignard}, {Murray}, {Le Poole}, {Schrijver}, {Turon},
  {Arenou}, {Froeschl{\'e}}, \& {Petersen}}]{1997A&A...323L..49P}
{Perryman}, M.~A.~C., {Lindegren}, L., {Kovalevsky}, J., {Hoeg}, E., {Bastian},
  U., {Bernacca}, P.~L., {Cr{\'e}z{\'e}}, M., {Donati}, F., {Grenon}, M., {van
  Leeuwen}, F., {van der Marel}, H., {Mignard}, F., {Murray}, C.~A., {Le
  Poole}, R.~S., {Schrijver}, H., {Turon}, C., {Arenou}, F., {Froeschl{\'e}},
  M., \& {Petersen}, C.~S. 1997, \aap, 323, L49

\bibitem[{{Riedel} et~al.(2010){Riedel}, {Subasavage}, {Finch}, {Jao}, {Henry},
  {Winters}, {Brown}, {Ianna}, {Costa}, \& {Mendez}}]{2010AJ....140..897R}
{Riedel}, A.~R., {Subasavage}, J.~P., {Finch}, C.~T., {Jao}, W., {Henry},
  T.~J., {Winters}, J.~G., {Brown}, M.~A., {Ianna}, P.~A., {Costa}, E., \&
  {Mendez}, R.~A. 2010, \aj, 140, 897. \eprint{1008.0648}

\bibitem[{{Samus} et~al.(2009){Samus}, {Durlevich}, \& {et
  al.}}]{2009yCat....102025S}
{Samus}, N.~N., {Durlevich}, O.~V., \& {et al.} 2009, VizieR Online Data
  Catalog, 1, 2025

\bibitem[{{Subasavage} et~al.(2005){Subasavage}, {Henry}, {Hambly}, {Brown}, \&
  {Jao}}]{2005AJ....129..413S}
{Subasavage}, J.~P., {Henry}, T.~J., {Hambly}, N.~C., {Brown}, M.~A., \& {Jao},
  W. 2005, \aj, 129, 413. \eprint{arXiv:astro-ph/0409505}

\bibitem[{{Voges} et~al.(1999){Voges}, {Aschenbach}, {Boller},
  {Br{\"a}uninger}, {Briel}, {Burkert}, {Dennerl}, {Englhauser}, {Gruber},
  {Haberl}, {Hartner}, {Hasinger}, {K{\"u}rster}, {Pfeffermann}, {Pietsch},
  {Predehl}, {Rosso}, {Schmitt}, {Tr{\"u}mper}, \&
  {Zimmermann}}]{1999A&A...349..389V}
{Voges}, W., {Aschenbach}, B., {Boller}, T., {Br{\"a}uninger}, H., {Briel}, U.,
  {Burkert}, W., {Dennerl}, K., {Englhauser}, J., {Gruber}, R., {Haberl}, F.,
  {Hartner}, G., {Hasinger}, G., {K{\"u}rster}, M., {Pfeffermann}, E.,
  {Pietsch}, W., {Predehl}, P., {Rosso}, C., {Schmitt}, J.~H.~M.~M.,
  {Tr{\"u}mper}, J., \& {Zimmermann}, H.~U. 1999, \aap, 349, 389.
  \eprint{arXiv:astro-ph/9909315}

\bibitem[{{Voges} et~al.(2000){Voges}, {Aschenbach}, {Boller}, {Brauninger},
  {Briel}, {Burkert}, {Dennerl}, {Englhauser}, {Gruber}, {Haberl}, {Hartner},
  {Hasinger}, {Pfeffermann}, {Pietsch}, {Predehl}, {Schmitt}, {Trumper}, \&
  {Zimmermann}}]{2000IAUC.7432R...1V}
{Voges}, W., {Aschenbach}, B., {Boller}, T., {Brauninger}, H., {Briel}, U.,
  {Burkert}, W., {Dennerl}, K., {Englhauser}, J., {Gruber}, R., {Haberl}, F.,
  {Hartner}, G., {Hasinger}, G., {Pfeffermann}, E., {Pietsch}, W., {Predehl},
  P., {Schmitt}, J., {Trumper}, J., \& {Zimmermann}, U. 2000, iaucirc, 7432, 1

\end{thebibliography}

\end{document}